\documentclass[twocolumn,ruled,vlined,conference]{IEEEtran}
\usepackage[latin9]{inputenc}
\usepackage{array}
\usepackage{verbatim}
\usepackage{algorithm2e}
\usepackage{amsmath}
\usepackage{graphicx}

\makeatletter

\providecommand{\tabularnewline}{\\}

\usepackage[final]{pdfpages}
\usepackage{cite}
\usepackage{amsmath}
\usepackage{amsfonts}
\usepackage{amssymb}
\usepackage{amsthm}
\usepackage{multirow}
\usepackage{array}
\usepackage{slashbox}
\usepackage{makecell}
\usepackage{threeparttable}
\usepackage{algorithmic}
\usepackage{array}
\newcolumntype{I}{!{\vrule width 2pt}}
\usepackage{makecell}% for auto line breaking
\usepackage{graphicx}

\usepackage{rotating}

\usepackage{pict2e}
\usepackage{cases}
\usepackage{enumitem}
\usepackage{footnote}
\makesavenoteenv{tabular}
\makesavenoteenv{table}

\usepackage{floatrow}
\usepackage{url}
\newcommand{\figref}{Figure }
\newcommand{\tabref}{Table }
\newcommand{\secref}{Section }

\title{On Path Memory in List Successive Cancellation Decoder of Polar Codes}

\author{ChenYang Xia, YouZhe Fan, Ji Chen, Chi-Ying Tsui\\ Department of Electronic and Computer Engineering, the HKUST, Hong Kong\\\{cxia, jasonfan, jchenbh\}@connect.ust.hk, eetsui@ust.hk\\}
\@ifundefined{showcaptionsetup}{}{%
 \PassOptionsToPackage{caption=false}{subfig}}
\usepackage{subfig}
\makeatother

\begin{document}
\maketitle
\begin{abstract}
Polar code is a breakthrough in coding theory. Using list successive
cancellation decoding with large list size $\mathcal{L}$, polar codes
can achieve excellent error correction performance. The $\mathcal{L}$
partial decoded vectors are stored in the path memory and updated
according to the results of list management. In the state-of-the-art
designs, the memories are implemented with registers and a large crossbar
is used for copying the partial decoded vectors from one block of
memory to another during the update. The architectures are quite area-costly
when the code length and list size are large. To solve this problem,
we propose two optimization schemes for the path memory in this work.
First, a folded path memory architecture is presented to reduce the
area cost. Second, we show a scheme that the path memory can be totally
removed from the architecture. Experimental results show that these
schemes effectively reduce the area of path memory.
\end{abstract}

\begin{IEEEkeywords}
Polar codes, List successive cancellation decoding, Path memory, Partial-sums
\end{IEEEkeywords}

\section{Introduction\label{sec:intro}}

Polar codes \cite{earikan_bilkent_tit_2009_polar} are the first kind
of forward error correction code that is proved to achieve channel
capacity. The basic decoding scheme of polar codes is called successive
cancellation decoding (SCD) \cite{earikan_bilkent_tit_2009_polar}.
The decoding is sequential in nature as the decoding of a new bit
has dependency on the already decoded bits. Specifically, the data
dependency comes from the partial-sums which are obtained by encoding
some of the already decoded bits. These partial-sums are used as the
inputs of the subsequent computations. List successive cancellation
decoding (LSCD), proposed in \cite{ital_ucsd_tit_2015_list}, includes
$\mathcal{L}$ parallelly-decoded SCDs and keeps $\mathcal{L}$ partial
decoded vectors during decoding. By using a large list size and selecting
a decoded vector satisfying the cyclic redundancy check (CRC) \cite{kniu_bupt_cl_2012_crc,bli_huawei_cl_2012_crc}
at the end of decoding, the error correction performance of polar
codes is greatly improved. 

The hardware architecture of LSCD \cite{abalatsoukas_epfl_tsp_2015_llrlscd,yzfan_hkust_jsac_2016_sedts,sahashemi_mcgill_tsp_2017_fastflexible,jlin_lehigh_tvlsi_2016_highthpt,crxiong_lehigh_tsp_2016_symbol,smabbas_hkust_sips_2015_bp}
implements $\mathcal{L}$ SCD kernels to support the parallel calculations
of $\mathcal{L}$ paths, indicating the hardware complexity is at
least $\mathcal{L}$ times that of the single SCD. To achieve a moderate
hardware complexity, the semi-parallel architecture \cite{cleroux_mcgill_tsp_2013_semiparallel}
and the folded partial-sum network (PSN) \cite{yzfan_hkust_tsp_2014_effps}
originally proposed for a single SCD are also adopted in the LSCD
architecture \cite{abalatsoukas_epfl_tsp_2015_llrlscd,yzfan_hkust_jsac_2016_sedts,sahashemi_mcgill_tsp_2017_fastflexible,jlin_lehigh_tvlsi_2016_highthpt,crxiong_lehigh_tsp_2016_symbol}
due to their low complexity. Several $\mathcal{L}\times\mathcal{L}$
crossbars are used for the permutations of log-likelihood ratios (LLRs),
partial-sums and partial decoded vectors among different memories
according to the results of list management. In \cite{abalatsoukas_epfl_tsp_2015_llrlscd},
pointers are used to access the corresponding LLRs for computation
instead of directly copying the LLRs during update.

The path memory with $\mathcal{L}\cdot N$ memory bits is implemented
to store the $\mathcal{L}$ partial decoded vectors in the existing
LSCD architecture \cite{yzfan_hkust_jsac_2016_sedts,abalatsoukas_epfl_tsp_2015_llrlscd,sahashemi_mcgill_tsp_2017_fastflexible}.
It updates the contents when a new bit is decoded. The partial decoded
vector of a path is duplicated if both its expanded paths are kept.
After this duplication, an $N$-bit crossbar is needed for the permutation
of the updated partial decoded vectors, which has a very high complexity
and takes a large area when the code length and list size are large.
Moreover, the path memory needs to be implemented with registers which
usually have a larger area than SRAMs.

In this work, to optimize the complexity of the path memory, we first
propose a folded path memory by mimicking the architecture of the
folded PSN. Then we present a method to recover the decoded bits from
the partial-sums which are already available in the folded PSN and
hence the path memory can be omitted. It is shown that the latency
of this recovery can be hidden in the decoding process, therefore
will not cause any latency overhead in most conditions. 

\begin{comment}
Although the information brought by the partial-sums and the decoded
bits are actually the same, they are stored in two different memories
in the existing LSCD architecture \cite{abalatsoukas_epfl_tsp_2015_llrlscd,yzfan_hkust_jsac_2016_sedts,sahashemi_mcgill_tsp_2017_fastflexible}.
Moreover, as a larger list size brings a better performance, 

The hardware architecture mapping of SCD has been studied by various
researchers. Most of these proposed architectures include two datapaths,
i.e., a feedforward datapath and a feedback datapath. The feedforward
datapath calculates the soft values, usually log-likelihood ratios
(LLRs), to decode the source bits, while the feedback datapath generates
the partial-sums. To get a higher hardware utilization, these two
datapaths are optimized in \cite{cleroux_mcgill_tsp_2013_semiparallel}
and \cite{yzfan_hkust_tsp_2014_effps}, respectively.
\end{comment}

\subsubsection*{Notation}

In this paper, matrices and row vectors are denoted in boldface uppercase
and lowercase letters, respectively. $\mathbf{X}_{M}$ represents
a square matrix of order $M$ and $\mathbf{x}_{M}$ represents an
$M$-dimensional vector. $x_{i}$ is the $i^{th}$ element of a vector
$\mathbf{x}$.

\section{Miscellaneous\label{sec:misc}}

\subsection{Introduction of Polar Codes\label{subsec:polar}}

\begin{figure}
\subfloat[]{\includegraphics{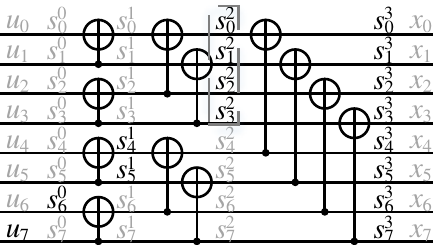}

}\subfloat[]{\includegraphics{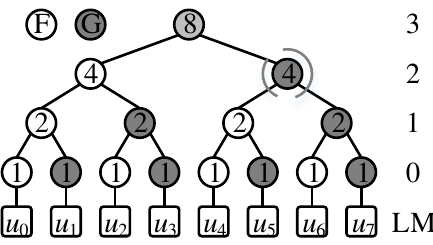}}

\subfloat[]{%
\begin{tabular}{c|c|c|c|c|c|c|c|c}
\hline 
Cycle & \multicolumn{2}{c|}{0} & \multicolumn{2}{c|}{1} & \multicolumn{2}{c|}{2} & \multicolumn{2}{c}{3}\tabularnewline
\hline 
\hline 
Input 0 & $s_{7}^{1}$ & $s_{6}^{1}$ & $s_{5}^{1}$ & $s_{4}^{1}$ & $s_{3}^{2}$ & $s_{2}^{2}$ & $s_{1}^{2}$ & $s_{0}^{2}$\tabularnewline
\hline 
Input 1 & - & - & $s_{7}^{3}$ & $s_{6}^{3}$ & $s_{7}^{3}$ & $s_{6}^{3}$ & $s_{5}^{3}$ & $s_{4}^{3}$\tabularnewline
\hline 
Output & $s_{7}^{3}$ & $s_{6}^{3}$ & $s_{5}^{3}$ & $s_{4}^{3}$ & $s_{3}^{3}$ & $s_{2}^{3}$ & $s_{1}^{3}$ & $s_{0}^{3}$\tabularnewline
\hline 
\end{tabular}

}\caption{(a) Encoding signal flow graph and (b) LSCD scheduling tree of polar
codes and (c) the steps of partial-sum update in a folded PSN ($\Lambda=8$
and $P=2$).}
\label{fig:codec}
\end{figure}

Polar codes \cite{earikan_bilkent_tit_2009_polar} are a kind of linear
block codes whose code length is denoted as $N$. Its generator matrix
is Kronecker matrix $\mathbf{F}^{\otimes n}$, where $\mathbf{F}=\left[\begin{array}{cc}
1 & 0\\
1 & 1
\end{array}\right]$ and $n=log_{2}N$. A codeword $\mathbf{x}_{N}$ can be encoded from
a source word $\mathbf{u}_{N}$ by $\mathbf{x}_{N}=\mathbf{u}_{N}\cdot\mathbf{F}^{\otimes n}$,
where $\mathbf{u}_{N},\mathbf{x}_{N}\in\{0,1\}^{N}$. \figref \ref{fig:codec}(a)
shows an encoding signal flow graph of $\mathbf{F}^{\otimes3}$ in
which each ``$\oplus$'' node executes an XOR operation and each
``$\cdot$'' node split its input%
\begin{comment}
in which $\mathbf{u}_{8}$ is inputted from the left and $\mathbf{x}_{8}$
is outputted to the right
\end{comment}
. 

LSCD of polar codes can be represented by a scheduling tree shown
in \figref \ref{fig:codec}(b). It includes two parts. The upper
half is a full binary tree with $n+1$ stages, representing $\mathcal{L}$
identical SCDs for $\mathcal{L}$ paths. The stage indices are in
descending order from the root to the leaf nodes. Two kinds of nodes,
denoted as F-nodes and G-nodes, exist in this tree. The number of
functions executed in each node is also marked in \figref \ref{fig:codec}(b).
The functions in G-nodes depend on the partial-sums which are encoded
from the already decoded bits. Specifically, the partial-sums for
the $j^{th}$ node from the left on stage $\lambda$ are calculated
as
\begin{equation}
[\hat{s}_{(j-1)\cdot\Lambda}^{\lambda},...,\hat{s}_{j\cdot\Lambda-1}^{\lambda}]=[\hat{u}_{(j-1)\cdot\Lambda},...,\hat{u}_{j\cdot\Lambda-1}]\cdot\mathbf{F}^{\otimes\lambda},\label{eq:ps}
\end{equation}
where stage index $\lambda\in[0,n-1]$ and $j\in[0,2^{n-\lambda}-1]$,
$\Lambda=2^{\lambda}$ is the bit width of partial-sums at stage $\lambda$
and $\hat{u}$ and $\hat{s}$ are the decoded bits and partial-sums,
respectively\footnote{For simplicity, the hats in $\hat{u}$ and $\hat{s}$ are omitted
in the rest of this paper}. For example, the G-node at stage 2 $(j=1)$ in \figref \ref{fig:codec}(b)
needs the partial-sums generated from $[\hat{u}_{0},...,\hat{u}_{3}]$
in \figref \ref{fig:codec}(a).

The source bits are decoded in an ascending order. At each leaf node,
a source bit is decoded. For each path in LSCD, either possibility
that the decoded bit is 0 or 1 is considered and the number of paths
is doubled. If the number of paths exceeds the list size $\mathcal{L}$,
list management operations, denoted by the squares in the scheduling
tree, are executed to keep the best $\mathcal{L}$ decoding paths
in the list and discard the others.%
\begin{comment}
an original path will be expanded to two by keeping . A path metric
assigned to each path to represent its reliability is also updated
accordingly. The number of paths increases exponentially with respect
to the already decoded bits. In order to limit the decoding complexity,
list management operations, represented by the lower half square in
the scheduling tree, is used to find the best $\mathcal{L}$ decoding
paths. 
\end{comment}

\subsection{Folded Partial-sum Network\label{subsec:fpsn}}

\begin{figure}
\includegraphics{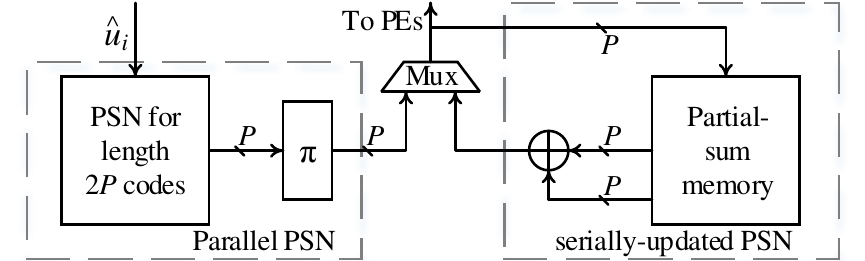}

\caption{The block diagram of folded partial-sum network.}
\label{fig:fpsn}
\end{figure}

In SCD, $2^{\lambda}$ F- or G-functions can be calculated in parallel
at stage $\lambda$, so $\frac{N}{2}$ processing elements (each is
used to calculate one function) should be implemented if we want to
maximize the parallelism. However, the area cost will be very high.
To reduce the hardware complexity, semi-parallel architecture \cite{cleroux_mcgill_tsp_2013_semiparallel}
was proposed to limit the computational parallelism to $P=2^{p}(\ll N)$,
i.e., at most $P$ functions are calculated in one clock cycle and
a node is calculated in $\left\lceil 2^{\lambda-p}\right\rceil $
clock cycles. The complexity of PSN in this kind of semi-parallel
architecture can also be reduced. In \cite{yzfan_hkust_tsp_2014_effps},
an folded PSN architecture is proposed, which generates at most $P$
partial-sum bits in one clock cycle. Its block diagram is shown in
\figref \ref{fig:fpsn}. The partial-sums for the nodes at stages
not higher than $p$ are updated by a parallel PSN and stored in a
$P$-bit register bank. The partial-sums at higher stages are serially
updated in a word of $P$ bits and stored in an $\frac{N}{2}$-bit
SRAM. The encoding signal flow graph in \figref \ref{fig:codec}(a)
can be used to illustrate how to generate $\Lambda=8$ bits of partial-sums
for G-node at stage 3. Supposing that the parallelism $P=2$, the
partial-sums already generated are $\{[s_{0}^{2},s_{1}^{2},s_{2}^{2},s_{3}^{2}],[s_{4}^{1},s_{5}^{1}],s_{6}^{0}\}$
and the newly decoded bits is $u_{7}$. First, $[s_{6}^{1},s_{7}^{1}]$
are parallelly updated from $s_{6}^{0}$ and $u_{7}$. Then, the required
partial-sums $[s_{0}^{3},...,s_{7}^{3}]$ are generated according
to the schedule shown in \figref \ref{fig:codec}(c) within 4 clock
cycles. According to the synthesis results in \cite{yzfan_hkust_tsp_2014_effps},
the folded PSN has a much smaller area than other fully-parallel PSN
architecture, such as the partial-sum update logic \cite{cleroux_mcgill_tsp_2013_semiparallel}
and the feed forward architecture \cite{czhang_umn_tsp_2013_overlap}. 

\subsection{Problems of the Existing Path Memory\label{subsec:existing}}

\begin{figure}
\includegraphics{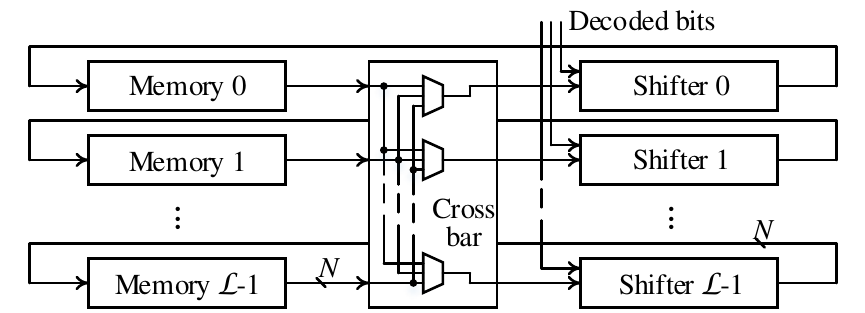}

\caption{The block diagram of the traditional path memory.}
\label{fig:path_mem}
\end{figure}

\begin{comment}
In an LSCD, $\mathcal{L}$ traditonal SCDs are implemented. Besides,
a path memory is implemented to store the partial decoded vectors. 
\end{comment}
The block diagram of the traditional path memory architecture for
LSCD \cite{yzfan_hkust_jsac_2016_sedts,abalatsoukas_epfl_tsp_2015_llrlscd,sahashemi_mcgill_tsp_2017_fastflexible}
is shown in \figref \ref{fig:path_mem}. $\mathcal{L}$ blocks of
memories are implemented to store the partial decoded vectors of $\mathcal{L}$
paths. Each memory includes $N$ bits of registers. After the list
management operation is executed, some paths are pruned while other
paths are kept and duplicated, and the contents in the path memory
are updated. First, the crossbar permutes the paths according to the
list management results. Then the newly decoded bit of each path is
appended to the corresponding permuted partial decoded vector by a
shifter. Finally, the updated paths are stored in the path memory.
According to the synthesis results, the crossbar used in this architecture,
which has a quadratic complexity with respect to the list size, takes
a very large area when large code length and list size are used and
this becomes a significant issue of the existing architecture. The
registers also take a large area and are expected to be substituted
with other hardware-friendly memory elements.

\section{Folded Path Memory\label{sec:fpm}}

\begin{figure}
\includegraphics{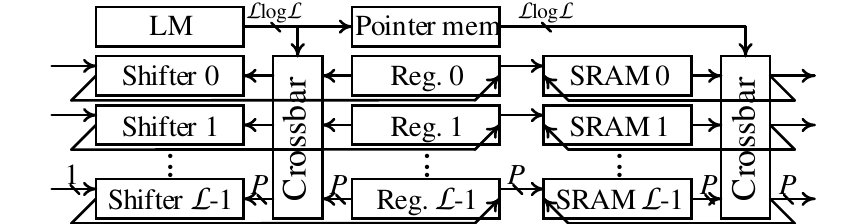}

\caption{The block diagram of the proposed folded path memory.}
\label{fig:fpm}
\end{figure}
\begin{figure}
\subfloat[]{\includegraphics{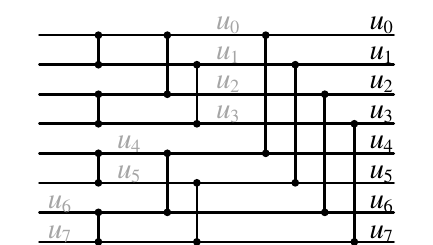}}\subfloat[]{\includegraphics{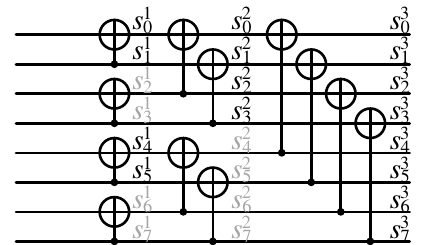}

}

\subfloat[]{%
\begin{tabular}{c|c|c|c|c|c|c|c|c}
\hline 
Cycle & \multicolumn{2}{c|}{0} & \multicolumn{2}{c|}{1} & \multicolumn{2}{c|}{2} & \multicolumn{2}{c}{3}\tabularnewline
\hline 
\hline 
Input 0 & $s_{0}^{3}$ & $s_{1}^{3}$ & $s_{2}^{3}$ & $s_{3}^{3}$ & $s_{0}^{2}$ & $s_{1}^{2}$ & $s_{4}^{2}$ & $s_{5}^{2}$\tabularnewline
\hline 
Input 1 & $s_{4}^{3}$ & $s_{5}^{3}$ & $s_{6}^{3}$ & $s_{7}^{3}$ & $s_{2}^{2}$ & $s_{3}^{2}$ & $s_{6}^{2}$ & $s_{7}^{2}$\tabularnewline
\hline 
``$\oplus$'' & $s_{0}^{2}$ & $s_{1}^{2}$ & $s_{2}^{2}$ & $s_{3}^{2}$ & $s_{0}^{1}$ & $s_{1}^{1}$ & $s_{4}^{1}$ & $s_{5}^{1}$\tabularnewline
\hline 
``$\cdot$'' & $s_{4}^{2}$ & $s_{5}^{2}$ & $s_{6}^{2}$ & $s_{7}^{2}$ & $s_{2}^{1}$ & $s_{3}^{1}$ & $s_{6}^{1}$ & $s_{7}^{1}$\tabularnewline
\hline 
\end{tabular}

}\caption{The signal flow graph of (a) folded path memory and (b) decoded bits
recovery and (c) the recovery schedule ($\Lambda=8$ and $P=2$). }

\label{fig:sfg_pps}
\end{figure}

As discussed in \secref \ref{subsec:existing}, in the traditional
path memory, the area overhead is mainly due to the $N$-bit crossbar
when the list size is large.%
\begin{comment}
 is implemented as the $\mathcal{L}$ partial decoded vectors need
to be permuted according to the list management results, which is
\end{comment}
{} Consequently, the key to reduce the complexity of the path memory
is to reduce the crossbar size. As presented in \secref \ref{subsec:fpsn},
the folded PSN updates at most $P$ partial-sum bits in one clock
cycle. If this architecture is used in an LSCD, the crossbar size
is only $P$ bits, which is much smaller than that of the $N$-bit
crossbar in a parallel path memory. According to \secref \ref{subsec:fpsn},
the partial-sums and decoded bits have the same bit width and are
always updated at the same time during the decoding. Based on these
observations, we propose an architecture called folded path memory
which mimics the architecture of the folded PSN, as shown in \figref
\ref{fig:fpm}. 

The left part of the folded path memory includes $\mathcal{L}$ $P$-bit
register banks, $\mathcal{L}$ shifters and a $P$-bit crossbar. After
the list management operation, the crossbar read the $P$-bit partial
decoded vectors from the register banks and update them in the same
way as the parallel path memory shown in \figref \ref{fig:path_mem}.
\begin{comment}
and permute them according to the list management results. Then each
newly decoded bit is appended to the permuted path. Finally the updated
partial decoded vectors are stored back to the register banks. 
\end{comment}
When each register bank in the left part is full with $P$ bits, these
bits are sent to the right part.

The right part uses $\mathcal{L}$ blocks of SRAMs to store the partial
decoded vectors. The port width of each SRAM is $P$ bits and its
total size equals to $N$ bits. The stored vectors are not permuted
for update. Instead, we can use $\frac{N}{P}$ pointers to store the
block indices of the SRAM in which each $P$ bits are stored. However,
to update the pointers, we still need extra hardware. To use as few
pointers as possible, we still use a crossbar to permute the decoded
bits which have the same indices with the partial-sums that are being
updated. Take an example with $\Lambda=8$ and $P=2$, whose signal
flow graph shown in \figref \ref{fig:sfg_pps}(a) can be obtained
by changing the ``$\oplus$'' nodes in \figref \ref{fig:codec}(a)
to ``$\cdot$'' nodes. During the four clock cycles when the partial-sums
at stage 3, $[s_{0}^{3},...,s_{7}^{3}]$, are generated, the corresponding
$[u_{0},...,u_{3}]$, $[u_{4},u_{5}]$ and $[u_{6},u_{7}]$ of this
path but previously stored in different blocks of memories are permuted
through the crossbar and stored in the SRAM of this path in these
four cycles. By doing so, $[u_{0},...,u_{7}]$ of each path can be
pointed by a pointer instead of three pointers. For a polar code with
code length equal to $N$, the partial decoded bits are store in $n-p+1$
groups with their length $\Lambda\in\{\frac{N}{2},\frac{N}{4},...,2P,P,P\}$.
This means only $n-p+1$ pointers are enough for each path.

 %
\begin{comment}
of both the read port and write port of an SRAM is $P$ bits. 
\end{comment}

Finally, as the two crossbars are never activated simultaneously,
only one crossbar is implemented in the final architecture. Comparing
with the existing architectures, the folded path memory uses a much
smaller crossbar while it is adaptive to LSCD with any code length
and list size and also easy to implement. %
\begin{comment}
explain permute

(do not need to introduce architecture list FPSN)

(recover partial-sum: already stored in the partial-sum memory)
\end{comment}

\section{Recovering Decoded Bits from Partial-sums}

\begin{figure}
\includegraphics{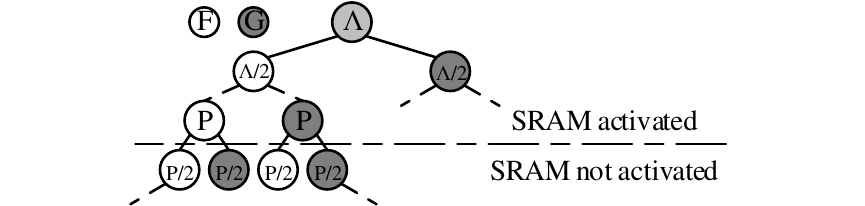}

\caption{The scheduling tree of recovering the decoded bits.}
\label{fig:scheduling-tree-lambda}
\end{figure}

In this section, based on the fact that the partial-sums are encoded
from the decoded bits, we introduce a scheme to directly recover the
decoded bits from the partial-sums stored in a folded PSN. We also
show that the proposed scheme do not introduce any extra latency comparing
with the traditional semi-parallel decoding schedule. By doing so,
the folded PSN and the path memory are merged and the path memory
can be omitted.

We rewrite \eqref{eq:ps} in the form of block matrices, i.e., we
divide the vectors into $P$-dimensional sub-vectors and the generator
matrix into sub-matrices of order $P$ and we have
\begin{align}
[(\mathbf{s}_{P}^{\lambda})_{0},...,(\mathbf{s}_{P}^{\lambda})_{\frac{\Lambda}{P}-1}] & =\nonumber \\{}
[(\mathbf{u}_{P})_{0},...,(\mathbf{u}_{P})_{\frac{\Lambda}{P}-1}] & \cdot(\mathbf{F}^{\otimes p}\otimes\mathbf{F}^{\otimes\lambda-p})\label{eq:dividing}
\end{align}
where $(\mathbf{s}_{P}^{\lambda})_{j}=[s_{j\cdot P}^{\lambda},...,s_{(j+1)\cdot P-1}^{\lambda}]$
and $(\mathbf{u}_{P})_{j}=[u_{j\cdot P},...,u_{(j+1)\cdot P-1}]$
$(j\in[0,\frac{\Lambda}{P}-1])$. Each $P$-bit sub-vector $(\mathbf{s}_{P}^{\lambda})_{j}$
is the content stored in one address in the SRAM of the folded PSN
and is a linear combination of $(\mathbf{u}_{P})_{j}\cdot\mathbf{F}^{\otimes p}$.
Consequently, to recover the decoded bits from the partial-sums, we
first calculate all the intermediate values $(\mathbf{u}_{P})_{j}\cdot\mathbf{F}^{\otimes p}$
from the partial-sums, then we encode the intermediate values to get
the corresponding decoded bits because $(\mathbf{u}_{P})_{j}\cdot\mathbf{F}^{\otimes p}\cdot\mathbf{F}^{\otimes p}=(\mathbf{u}_{P})_{j}\cdot\mathbf{I}_{P}=(\mathbf{u}_{P})_{j}$,
where $\mathbf{I}$ is an identity matrix.

Next, we derive the equations of $(\mathbf{u}_{P})_{j}\cdot\mathbf{F}^{\otimes p}$.
From the mixed-product property\footnote{$(\mathbf{A}\otimes\mathbf{B})(\mathbf{C}\otimes\mathbf{D})=(\mathbf{AC})\otimes(\mathbf{BD})$,
if $\mathbf{AC}$ and $\mathbf{BD}$ exist.} of Kronecker product, we can get
\begin{equation}
(\mathbf{F}^{\otimes p}\otimes\mathbf{F}^{\otimes\lambda-p})\cdot(\mathbf{I}_{P}\otimes\mathbf{F}^{\otimes\lambda-p})=\mathbf{F}^{\otimes p}\otimes\mathbf{I}_{\frac{\Lambda}{P}}.\label{eq:mixed-product}
\end{equation}
Multiply both sides of \eqref{eq:dividing} by $(\mathbf{I}_{P}\otimes\mathbf{F}^{\otimes\lambda-p})$,
we can get
\begin{align}
[(\mathbf{s}_{P}^{\lambda})_{0},...,(\mathbf{s}_{P}^{\lambda})_{\frac{\Lambda}{P}-1}] & \cdot(\mathbf{I}_{P}\otimes\mathbf{F}^{\otimes\lambda-p})=\nonumber \\{}
[(\mathbf{u}_{P})_{0},...,(\mathbf{u}_{P})_{\frac{\Lambda}{P}-1}] & \cdot(\mathbf{F}^{\otimes p}\otimes\mathbf{I}_{\frac{\Lambda}{P}}).\label{eq:dot-product}
\end{align}
Take an numerical example of \eqref{eq:dot-product} with $\frac{\Lambda}{P}=4$,
by using the multiplication of block matrices, we can get
\begin{equation}
\begin{cases}
(\mathbf{u}_{P})_{0}\cdot\mathbf{F}^{\otimes p} & =(\mathbf{s}_{P}^{\lambda})_{0}+(\mathbf{s}_{P}^{\lambda})_{1}+(\mathbf{s}_{P}^{\lambda})_{2}+(\mathbf{s}_{P}^{\lambda})_{3}\\
(\mathbf{u}_{P})_{1}\cdot\mathbf{F}^{\otimes p} & =(\mathbf{s}_{P}^{\lambda})_{1}+(\mathbf{s}_{P}^{\lambda})_{3}\\
(\mathbf{u}_{P})_{2}\cdot\mathbf{F}^{\otimes p} & =(\mathbf{s}_{P}^{\lambda})_{2}+(\mathbf{s}_{P}^{\lambda})_{3}\\
(\mathbf{u}_{P})_{3}\cdot\mathbf{F}^{\otimes p} & =(\mathbf{s}_{P}^{\lambda})_{3}
\end{cases}\label{eq:dot-product-4}
\end{equation}
\begin{comment}
$[(\mathbf{u}_{P})_{0},(\mathbf{u}_{P})_{1},(\mathbf{u}_{P})_{2},(\mathbf{u}_{P})_{3}]=[(\mathbf{s}_{P})_{0},(\mathbf{s}_{P})_{1},(\mathbf{s}_{P})_{2},(\mathbf{s}_{P})_{3}]\cdot\left[\begin{array}{cccc}
\mathbf{F}^{\otimes p} & \mathbf{O} & \mathbf{O} & \mathbf{O}\\
\mathbf{F}^{\otimes p} & \mathbf{F}^{\otimes p} & \mathbf{O} & \mathbf{O}\\
\mathbf{F}^{\otimes p} & \mathbf{O} & \mathbf{F}^{\otimes p} & \mathbf{O}\\
\mathbf{F}^{\otimes p} & \mathbf{F}^{\otimes p} & \mathbf{F}^{\otimes p} & \mathbf{F}^{\otimes p}
\end{array}\right]$$\{\mathbf{s}_{N/2}^{n-1},\mathbf{s}_{N/4}^{n-2},...,\mathbf{s}_{P}^{p},\mathbf{s}_{P}^{p}\}$$\{\Lambda_{i}|\Lambda_{i}\in[\frac{N}{2},\frac{N}{4},...,2P,P,P],i\in[0,n-p]\}.$
\end{comment}
The left hand side are the $(\mathbf{u}_{P})_{j}\cdot\mathbf{F}^{\otimes p}$
we want to calculate, and the right hand side are the $P$-bit sub-vectors
of the partial-sums. So \eqref{eq:dot-product} can be regarded as
encoding $P$ groups of $\frac{\Lambda}{P}$-bit sub-codes. With the
$P$ XOR gates in the folded PSN, one XOR calculation in each of the
$P$ groups of encoding is executed in one clock cycle. This indicates
that for a sub-code whose length equals to $\Lambda$, its latency
for recovery equals to the number of the ``$\oplus$'' nodes in
the encoding signal flow graph of a $\frac{\Lambda}{P}$-bit polar
code. Consequently, the latency to recover $\Lambda$ decoded bits
from the corresponding partial-sums is $\frac{\Lambda}{2P}log_{2}\frac{\Lambda}{P}$
clock cycles. For example, to recover the $\Lambda=8$ decoded bits
in an LSCD with $P=2$ in \figref \ref{fig:sfg_pps}(b), we encode
two 4-bit sub-codes, $[s_{0}^{3},s_{2}^{3},s_{4}^{3},s_{6}^{3}]$
and $[s_{1}^{3},s_{3}^{3},s_{5}^{3},s_{7}^{3}]$, whose schedule is
shown in \figref \ref{fig:sfg_pps}(c) and the total latency is 4
clock cycles. Finally, an extra $P$-bit encoder for each path is
used to encode $(\mathbf{u}_{P})_{j}\cdot\mathbf{F}^{\otimes p}$.

For an $N$-bit polar code, $n-p+1$ groups of partial decoded bits
with their length $\Lambda\in\{\frac{N}{2},\frac{N}{4},...,2P,P,P\}$
need to be recovered from the folded PSN. The size of the SRAM is
$N$ bits, which is twice that of a traditional folded PSN as the
$\frac{N}{2}$-bit partial-sums for stage $n-1$ are not stored in
a traditional folded PSN \cite{yzfan_hkust_tsp_2014_effps}. The corresponding
decoded bits can be recovered after the right most G-node at stage
$\lambda$ is calculated because these memory bits are never used
to store or update partial-sums in the subsequent decoding.%
\begin{comment}
An issue is that it takes some extra latency to recover the decoded
bits after the partial-sum is generated. 
\end{comment}
{} By using the cycles in which the folded PSN is idle, the latency
can be hidden in the decoding process. Specifically, as shown in \figref
\ref{eq:lat-drv}, all the clock cycles used to calculate the nodes
below stage $p$ before the beginning of the next recovery of $\frac{\Lambda}{2}$
bits can be used to recover the $\Lambda$ bits because the SRAMs
in the folded PSN are not activated. By calculation, the number of
cycles in these stages is $\Lambda(1-\frac{1}{P})$ and it should
be larger than the latency for the recovery of $\Lambda$ decoded
bits which is $\frac{\Lambda}{2P}log_{2}\frac{\Lambda}{P}$ cycles.
Thus, the relationship between $\Lambda$ and $P$ where no extra
latency is introduced can be derived as
\begin{equation}
\Lambda<P\cdot2^{2P-2}.\label{eq:lat-drv}
\end{equation}
With practical parallelism $P=64$ which is used in most of the existing
architecture \cite{yzfan_hkust_jsac_2016_sedts,abalatsoukas_epfl_tsp_2015_llrlscd,sahashemi_mcgill_tsp_2017_fastflexible,crxiong_lehigh_tsp_2016_symbol,jlin_lehigh_tvlsi_2016_highthpt},
\eqref{eq:lat-drv} is satisfied for the LSCD with code length even
equal to $N=2^{20}$. %
\begin{comment}
Even we generate only $P=8$ partial-sums per clock cycle, the method
can be used for polar codes up to $N=2\Lambda=2^{18}$.
\end{comment}

For simplicity, we call the folded PSN which can recover the decoded
bits the merged memory.%
\begin{comment}
while the $P$ is much smaller than the practical one and will cause
a large latency overhead according to \cite{cleroux_mcgill_tsp_2013_semiparallel}.

Let $N$ be the code length of a block of polar code. The quantization
of LLRs is denoted as $Q$. 

The memories in the polar code decoders are mainly used to store the
channel LLRs, the intermediate LLRs, the partial-sums and the decoded
bits. As shown in \tabref \ref{tab:usage_saving}, in an SCD \cite{cleroux_mcgill_tsp_2013_semiparallel},
the memory usage is $NQ$ bits for each of the two kinds of LLRs and
$N$ bits for each of the partial-sums and the decoded bits. In an
LSCD \cite{abalatsoukas_epfl_tsp_2015_llrlscd,yzfan_hkust_jsac_2016_sedts},
all the memories except the one for channel LLRs are duplicated by
$\mathcal{L}$ times to execute the calculations of all the paths
in parallel. 
\begin{table}
\caption{The memory usage and the saving achieved by the proposed scheme}
\label{tab:usage_saving}

\begin{tabular}{|c|c|c|}
\hline 
 & SCD & LSCD\tabularnewline
\hline 
Channel LLRs & $NQ$ & $NQ$\tabularnewline
\hline 
Intermediate LLRs & $NQ$ & $\mathcal{L}NQ$\tabularnewline
\hline 
Partial-sums & $N$ & $\mathcal{L}N$\tabularnewline
\hline 
Decoded bits & $N$ & $\mathcal{L}N$\tabularnewline
\hline 
Total & $N(2Q+2)$ & $N(\mathcal{L}Q+Q+2\mathcal{L})$\tabularnewline
\hline 
Saving & $\frac{1}{(2Q+2)}$ & $\frac{1}{((1+\frac{1}{\mathcal{L}})Q+2)}$\tabularnewline
\hline 
\end{tabular}
\end{table}
\end{comment}

\section{Implementation Results}

\begin{table}
\caption{The SRAM size in all the mentioned architectures}
\label{tab:sram}

\begin{tabular}{c|c|c}
\hline 
 & {\small{}Port width} & {\small{}SRAM size}\tabularnewline
\hline 
{\small{}Folded PSN} & {\small{}$2P$} & {\small{}$\frac{N}{2}$}\tabularnewline
\hline 
{\small{}Folded path memory} & {\small{}$P$} & {\small{}$N$}\tabularnewline
\hline 
{\small{}Merged memory} & {\small{}$2P$} & {\small{}$N$}\tabularnewline
\hline 
\end{tabular}
\end{table}
To show the area saving achieved by the proposed path memory architectures,
we synthesize folded PSN, traditional path memory, folded path memory
and merged memory in the LSCD with different combinations of list
size and code length with UMC 90 nm technology. The timing constraint
for all the designs is 1ns and $P=64$ for a fair comparison. All
the SRAMs used in these architectures are summarized in \tabref \ref{tab:sram}.
All of them have two ports so that they can read and write data at
the same time. The synthesis results are shown in \tabref \ref{tab:results}.%
\begin{comment}
 The port width of folded PSN and merged memory are $2P$ according
to \cite{yzfan_hkust_tsp_2014_effps}. 
\end{comment}
{} The area of pointer memory is not included as the pointers for LLR
memory are valid and can be reused for these memories. 

Comparing with the traditional path memory, the folded path memory
achieves an area saving of more than 50\% for all different list size
and code length combinations. It also has a smaller area than the
folded PSN and the merged memory as the read port width of the SRAM
is $P$ bits instead of $2P$ bits. 

For the merged memory, all the combinations satisfy \eqref{eq:lat-drv},
indicating the decoded bits can be recovered without any latency overhead
comparing with the traditional schedule. Each merged memory is slightly
larger than its corresponding folded PSN because of the extra encoders
and SRAM bits.

For the storage of both partial-sums and decoded bits, we can use
either a folded PSN and a folded path memory (``(1)+(3)'') or just
a merged memory (``Only (4)''). The sum of the area of a folded
PSN and a traditional path memory (``(1)+(2)'') is used as a benchmark
for comparison. The larger the list size and the code length are,
the more saving we can get from the proposed architecture. The merged
memory brings us the most saving as the path memory is no more needed.
Regarding to the area saving with respect to the whole LSCD, the area
of an LSCD with $N=2^{10}$ and $\mathcal{L}=16$ is 7.47 $mm^{2}$
according to \cite{yzfan_hkust_jsac_2016_sedts}, which means about
20\% of the total area can be saved if a merged memory is used in
an LSCD.
\begin{table}
\caption{The synthesis results with UMC 90nm technology (unit: $mm^{2}$)}
\label{tab:results}

{\small{}}%
\begin{tabular}{>{\centering}m{2.1cm}|>{\centering}p{0.62cm}|>{\centering}p{0.62cm}|>{\centering}p{0.62cm}|>{\centering}p{0.62cm}|>{\centering}p{0.62cm}|>{\centering}p{0.62cm}}
\hline 
{\small{}Code length} & \multicolumn{3}{c|}{{\small{}$2^{10}$}} & \multicolumn{3}{c}{{\small{}$2^{13}$}}\tabularnewline
\hline 
{\small{}List size} & {\small{}8} & {\small{}16} & {\small{}32} & {\small{}8} & {\small{}16} & {\small{}32}\tabularnewline
\hline 
\hline 
{\small{}(1) Folded PSN} & {\small{}0.416} & {\small{}0.894} & {\small{}2.018} & {\small{}0.826} & {\small{}1.713} & {\small{}3.655}\tabularnewline
\hline 
{\small{}(2) Traditional path memory} & {\small{}0.521} & {\small{}1.692} & {\small{}6.158} & {\small{}4.251} & {\small{}15.83} & {\small{}54.72}\tabularnewline
\hline 
{\small{}(3) Folded path memory} & {\small{}0.228} & {\small{}0.526} & {\small{}1.279} & {\small{}0.696} & {\small{}1.462} & {\small{}3.151}\tabularnewline
\hline 
{\small{}(4) Merged memory} & {\small{}0.511} & {\small{}1.056} & {\small{}2.288} & {\small{}1.165} & {\small{}2.365} & {\small{}4.905}\tabularnewline
\hline 
\hline 
{\small{}(1) + (2)} & {\small{}0.937} & {\small{}2.586} & {\small{}8.176} & {\small{}5.077} & {\small{}17.54} & {\small{}58.38}\tabularnewline
\hline 
{\small{}(1) + (3)} & {\small{}0.644} & {\small{}1.420} & {\small{}3.297} & {\small{}1.522} & {\small{}3.175} & {\small{}6.806}\tabularnewline
\hline 
{\small{}Only (4)} & {\small{}0.511} & {\small{}1.056} & {\small{}2.288} & {\small{}1.165} & {\small{}2.365} & {\small{}4.905}\tabularnewline
\hline 
\end{tabular}{\small \par}
\end{table}

\section{Conclusion}

In this paper, we propose two methods to optimize the hardware complexity
of the path memory in the LSCD of polar codes. The folded path memory
mimics the architecture of the folded PSN to reduce the bit width
of the crossbar. It is easy to implement and can be used in any semi-parallel
LSCD architecture. The merged memory can recover the decoded bits
from the partial-sums stored in the folded PSN in almost all the practical
LSCD and hence the path memory can be omitted. Synthesis results show
that a large area saving can be achieved.

\end{document}